\documentclass[12pt]{article}
\usepackage{amssymb}

\usepackage{graphicx}
\usepackage{amsmath}

\newtheorem{theorem}{Theorem}

\newtheorem{claim}[theorem]{Claim}

\input{tcilatex}

\begin{document}

\bigskip \baselineskip16pt

\begin{titlepage}

\begin{flushright}
hep-th/9812184
\end{flushright}
\vspace{16 mm}

\begin{center}
{\Large \bf Matrix Representations of Holomorphic Curves on $T_{4}$}

\vspace{3mm}

\end{center}

\vspace{8 mm}

\begin{center}

Lorenzo Cornalba

\vspace{3mm}

{\small \sl Center for Theoretical Physics
\footnote{Visiting from: Princeton University, Department of Physics,
Princeton, NJ 08544, U.S.A.; cornalba@princeton.edu.}} \\
{\small \sl Massachusetts Institute of Technology} \\
{\small \sl Cambridge, MA 02139, U.S.A.} \\
{\small \tt cornalba@ctp.mit.edu}

\end{center}

\vspace{8mm}

\begin{abstract}

We construct a matrix representation of compact membranes analytically
embedded in complex tori. Brane configurations give rise, via Bergman
quantization, to $U(N)$ gauge fields on the dual torus, with almost-anti-self-dual
field strength. The corresponding $U(N)$ principal bundles are shown to be
non-trivial, with vanishing instanton number and first Chern class corresponding
to the homology class of the membrane embedded in the original torus. In the course
of the investigation, we show that the proposed quantization scheme naturally
provides an associative star-product over the space of functions on the surface,
for which we give an explicit and coordinate-invariant expression. This product
can, in turn, be used the quantize, in the sense of deformation quantization, any
symplectic manifold of dimension two.

\end{abstract}

\vspace{8mm}
\begin{flushleft}
Dicember 1998
\end{flushleft}
\end{titlepage}
\newpage 

\section{\label{s1}\label{i}\label{sec1}Introduction}

Matrix theory \cite{BFSS}\cite{Susskind-DLCQ}\cite{Seiberg-DLCQ}\cite{sen}
\cite{Bilal-review}\cite{Banks-review}\cite{Susskind-review} is believed to
describe, in the limit of large $N$, the fundamental degrees of freedom of $M
$-theory. In fact, within the same theory, both fundamental particles and
extended objects are described in a unified way. It is indeed remarkable
that one can start with a theory of gluons in a certain dimension (matrix
theory is nothing but $9+1$ $U(N)$ Super-Yang-Mills theory dimensionally
reduced to $0+1$ dimensions) and describe, in a dual way, a seemingly
unrelated theory of gravity in a different space-time dimension ($10+1$). In
light of this fact, it is important to better understand the relations that
exist between the two points of view, and to precisely describe how to
represent in matrix language objects which are familiar from the M-theory
prospective, and which are described at low energies within $11$-D
super-gravity.

It is of interest, in particular, to consider matrix theory configurations
which represent, within the gravity description, extended membranes. More
specifically one can study BPS membrane states, which only partially break
supersymmetry, and which are not expected, on general grounds, to be
effected by quantum corrections. Within the framework of supermembrane
theory \cite{Goldstone-Hoppe}\cite{dhn} , one can show that the BPS
condition is equivalent to the requirement that the brane be embedded
holomorphically in space, and it is therefore natural to look for matrix
representations of holomorphic curves.

In \cite{aaa} the question of representation of holomorphic curves embedded
in non-compact space was analyzed in detail. In particular the problem was
rephrased as a problem in geometric quantization, with $\varepsilon \sim
L_{P}^{3}/R$ playing the role of the Planck constant. A specific
quantization scheme was proposed, based on the concept of Bergman
projection, and the matrices representing the curve were taken to be
operators acting on the infinite dimensional space of holomorphic functions
living on the brane. In order to preserve the BPS character of the
configuration, one needs to choose a specific inner product on the space of
functions, which is related to a deformation of the K\"{a}hler potential of
the brane. Using an explicit expansion for the Bergman projection, the
deformation was determined asymptotically in $\varepsilon $.

In this paper we extend the results of \cite{aaa} to the interesting case of
holomorphic curves embedded in complex tori. The first major difference is
that the branes can now be taken to be compact. This requires an extension
of the quantization scheme proposed in \cite{aaa}, in which we take the
underlying Hilbert space to be the finite dimensional space of holomorphic
sections of a specific line bundle over the brane. The second major
difference with the basic case considered in \cite{aaa} comes from the fact
that, although the target space is still flat, we cannot quantize directly
the coordinate functions, since they are multivalued on the membrane. We
solve this problem using an extension of T-duality \cite{WT-Trieste}
appropriate to the present context, and we relate the brane configurations
on the torus to $U(N)$ Yang-Mills configurations on the dual torus. The
resulting $U(N)$ bundle is non-trivial, even though it has vanishing
instanton number. In fact the first Chern class of the bundle corresponds to
the homology class of the membrane embedded in the original torus. Moreover
the BPS character of the original membrane configuration is now translated
in a dual condition for the corresponding $U(N)$ gauge potential. More
precisely, we show that the corresponding field strength $\mathcal{F}$ is
almost-anti-self-dual, in the sense that 
\begin{eqnarray*}
\mathcal{F}_{12}+\mathcal{F}_{34} &\sim &\varepsilon \\
\mathcal{F}_{13}-\mathcal{F}_{24} &=&\mathcal{F}_{14}+\mathcal{F}_{23}=0.
\end{eqnarray*}
We also show that the configurations described are stable for to topological
reasons.

The basis of most of the discussion in this paper is the quantization scheme
which is analyzed in detail in the first part of the paper. In particular a
very important tool which is analyzed at length and used repeatedly, is a
specific non-commutative product (called star-product and denoted with $%
\star $) between functions on the brane. We show that the product $\star $,
which was introduced in \cite{aaa}, is an associative operation. In
particular, if we recast our result in the language of deformation
quantization \cite{bayen}\cite{lecomte}\cite{fedosov}\cite{maxim}\cite
{guillemin}, we show that the formula for the star-product can be used to
quantize any symplectic manifold of real dimension $2$.

The structure of this paper is as follows: in section \ref{bs} we review the
results of \cite{aaa} and extend them to a more general setting, which is
needed in the subsequent part of the discussion. In particular we introduce
the general concept of quantization used throughout the paper, and we define
a star product $\star $ on the space of functions. Section \ref{a} is
entirely devoted to show that the product $\star $ is actually associative.
This fact is heavily used in sections \ref{k} and \ref{t}, and connects our
discussion to the theory of deformation quantization. Section \ref{sb}
briefly describes how to go from a know star-product to a full quantization
scheme, and this result is then used in section \ref{k} to discuss general
results on the quantization of holomorphic curves embedded in K\"{a}hler
manifolds. Section \ref{tr} describes how to compute, asymptotically in the
quantization parameter $\varepsilon $, traces of operators obtained using
the quantization scheme described in section \ref{bs}. Finally section \ref
{t} is devoted to the main result of this paper. Using the results of the
previous sections, we show that we can associate to each holomorphic curve
embedded in a complex torus an almost-anti-self-dual Yang-Mills
configuration on the dual torus. We conclude the paper in section \ref{c}
with suggestions for future research.

\section{\label{bs}From Bergman Projections to Star Products}

In this first section we are going to recall and extend some of the main
results originally derived in \cite{aaa}. In particular we are going to
extend the results on Bergman projections and Bergman quantization \cite
{bergman}\cite{aaa}\cite{bordemann} to a more general setting, which is
needed to tackle the problem of representation of holomorphic curves
embedded in compact spaces. Although the exposition of the main ideas will
be self-contained, we will omit most of the proofs, since they are
essentially identical to the ones examined in \cite{aaa}. In what follows,
we will try to adhere, as much as possible, to the notation of \cite{aaa}.

Let us consider a compact Riemann surface $\Sigma $ of genus $g$, on which
we fix an arbitrary holomorphic line bundle $S$. We will denote by $K$ the
canonical line bundle on $\Sigma $, and by $T$ the product bundle 
\begin{equation*}
T=S^{-1}\otimes K.
\end{equation*}
Given two sections $\phi ,\psi $ of $S$, we wish to define an inner product $%
\left\langle \phi |\psi \right\rangle $ by integrating $\overline{\phi }\psi 
$ on the surface $\Sigma $ against a suitable measure. In order to do this
covariantly, we need to fix, first of all, a real and positive section $C$
of the line bundle $T\otimes \overline{T}$ \footnote{%
Note that the conditions of reality and positivity of $C$ are well defined,
since the bundle $T\otimes \overline{T}$ has real and positive definite
transition functions.}. It is then clear that the measure 
\begin{equation*}
\Omega (z)=i\,C(z)\,\ dz\wedge d\overline{z}
\end{equation*}
transforms as a section of $S^{-1}\otimes \overline{S}^{-1}$, and that the
expression $\overline{\phi }\psi \,\Omega $ represents a well defined $2$%
-form on the surface $\Sigma $. We can then define the inner-product $%
\left\langle \phi |\psi \right\rangle $ as 
\begin{equation*}
\left\langle \phi |\psi \right\rangle =\int_{\Sigma }\overline{\phi }\psi
\,\Omega .
\end{equation*}
Following the notation of \cite{aaa}, we will denote with $\mathcal{V}$ the
Hilbert space of sections of $S$ (not necessarily holomorphic) which have
finite norm with respect to the above inner product, and we shall call $%
\mathcal{H}\subset \mathcal{V}$ the subspace of $\mathcal{V}$ consisting of
holomorphic sections. Finally we will let $\pi $ be the orthogonal Bergman
projection onto $\mathcal{H}$ 
\begin{equation*}
\pi :\mathcal{V}\rightarrow \mathcal{H}.
\end{equation*}
Let us note that the choice of section $C$ not only provides us with a
specific inner product on $\mathcal{V}$, but also gives us a connection and
a covariant derivative for sections of specific line bundles. To be more
exact, let us introduce the connection 
\begin{equation*}
\Gamma =\partial \ln C
\end{equation*}
and let us consider a section $\phi $ of $T^{a}\otimes a.h.$, where $a.h.$
stands for an arbitrary anti-holomorphic line bundle. It is then easy to
show that 
\begin{equation*}
\nabla \phi =(\partial -a\Gamma )\phi
\end{equation*}
is a well-defined section of $K\otimes T^{a}\otimes a.h.$. A similar result
also holds for anti-holomorphic covariant derivatives. Finally we note that
the curvature of the connection just described is given by the $(1,1)$
tensor 
\begin{equation*}
R=\partial \overline{\partial }\ln C.
\end{equation*}

We now briefly summarize the properties of the projection $\pi $ in the
following
\newpage
\begin{claim}
The projection $\pi $ defined above satisfies the following properties

\begin{enumerate}
\item  The projection $\pi $ has an integral representation. More precisely,
let $h_{i}$ be an orthonormal basis for $\mathcal{H}$ and consider the
Bergman kernel 
\begin{equation*}
K(z,w)=\sum_{i}h_{i}(z)\overline{h}_{i}(w).
\end{equation*}
The kernel $K$ is a well-defined bi-section, independent of the choice of
orthonormal basis $h_{i}$ of $\mathcal{H}$. Moreover, if $\phi \in \mathcal{V%
}$, one has that 
\begin{equation*}
\pi (\phi )(z)=\int_{\Sigma }K(z,w)\phi (w)\,\Omega (w).
\end{equation*}

\item  If $\phi \in \mathcal{H}$ is holomorphic , then $\pi (\phi )=\phi .$

\item  For any $\phi \in \mathcal{V}$, one has that $\overline{\nabla }\pi
(\phi )=0.$

\item  Let $X$ be a section of $T^{-1}$ such that $\nabla X\in \mathcal{V}$
. Then $\pi (\nabla X)=0.$

\item  Let $X_{1},\cdots ,X_{n-1}$ be holomorphic $(-1,0)$ vector fields,
and let $X_{n}$ be a holomorphic section of $T^{-1}$. If $\phi $ is an
analytic function, then 
\begin{equation*}
\pi (\phi \nabla X_{1}\cdots \nabla X_{n})=(-1)^{n}X_{n}\cdots \nabla
X_{1}\nabla \phi .
\end{equation*}
\end{enumerate}
\end{claim}

The proof of the above claim is identical to the one given in \cite{aaa},
and we therefore refer the interested reader to that paper for further
details.

One of the main results of \cite{aaa} was to show that the projection $\pi $
possesses not only an integral representation but also, in an asymptotic
sense, a differential one. Again the argument is essentially identical in
the present setting, so that we content ourselves to state the following

\begin{claim}
Let $R=\partial \overline{\partial }\ln C$ be the curvature tensor.
Construct a sequence $P_{n}$ of $(1,1)$ tensors starting from 
\begin{equation*}
P_{1}=R
\end{equation*}
and using the recursion relation 
\begin{equation}
P_{n}=P_{n-1}+P_{1}+\partial \overline{\partial }\ln (P_{1}\cdots P_{n-1}).
\label{eq102}
\end{equation}
If $\phi \in \mathcal{V}$, we then have that 
\begin{equation*}
\pi (\phi )=\sum_{n=0}^{\infty }(-1)^{n}\nabla \frac{1}{P_{1}}\cdots \nabla 
\frac{1}{P_{n}}P_{1}\cdots P_{n}\frac{1}{P_{n}}\overline{\nabla }\cdots 
\frac{1}{P_{1}}\overline{\nabla }\phi ,
\end{equation*}
where 
\begin{equation*}
\nabla =\partial +\Gamma \,\ \ \ \ \ \ \ \ \ \ \ \ \overline{\nabla }=%
\overline{\partial }
\end{equation*}
\end{claim}

To conclude this section, we follow the general philosophy of \cite{aaa} and
discuss the process of quantization\cite{Woodhouse}. Consider a generic
function $A$ on the surface $\Sigma $, and let $\phi $ be an element of $%
\mathcal{H}$. The section $A\phi $, obtained by multiplying pointwise the
original section $\phi $ with the function $A$ is not necessarily
holomorphic. On the other hand we can extract the holomorphic part by acting
with the projection $\pi $, thus obtaining the element $\pi (A\phi )$ in $%
\mathcal{H}$. This process assigns to each function $A$ an operator on $%
\mathcal{H}$. More precisely, if we denote by $\frak{G}$ \ the space of
complex functions on $\Sigma $, we have constructed a quantization map $%
\mathcal{Q}:\frak{G}\rightarrow $End$(\mathcal{H})$ which associates to each
function $A$ the operator $\mathcal{A}=\mathcal{Q}(A)$ defined by 
\begin{equation*}
\mathcal{A}(\phi )=\pi (A\phi ).
\end{equation*}
The map $\mathcal{Q}$ is compatible with complex conjugation, since $%
\mathcal{Q}(\overline{A})=\mathcal{Q}^{\dagger }(A)$. Moreover it respects
the complex structure on $\Sigma $, since $\mathcal{Q}(A)\mathcal{Q}(B)=%
\mathcal{Q}(AB)$ whenever $A$ and $B$ are both holomorphic. Finally, using
the asymtotic expansion for the Bergman projection $\pi $ and following \cite
{aaa}, we can show that 
\begin{equation*}
\mathcal{Q}(A)\mathcal{Q}(B)=\mathcal{Q}(A\star B)
\end{equation*}
where we have introduced, on the space of functions $\frak{G}$, a product $%
\star $, called star product, given explicitly in terms of the tensors $%
P_{n} $ by the formula 
\begin{equation}
A\star B=\sum_{n=0}^{\infty }P_{1}\cdots P_{n}\left( \frac{1}{P_{n}}\partial
\cdots \frac{1}{P_{1}}\partial A\right) \left( \frac{1}{P_{n}}\overline{%
\partial }\cdots \frac{1}{P_{1}}\overline{\partial }B\right) .  \label{eq101}
\end{equation}

\section{\label{sec2}\label{a}Star Products: Proof of Associativity}

In section \ref{bs} we introduced a specific star product $\star $\ on the
space $\frak{G}$ of complex functions on the surface $\Sigma $. This section
is entirely devoted to show that the product $\star $ is associative.
\bigskip This fact will be repeadetly used in the subsequent sections, and
will be a key element in the basic construction of this paper described in
section \ref{t}. In particular are able to consider $\frak{G}$ as an
associative algebra, and we can therefore use most of our intuition about
operators on Hilbert spaces in the different context of functions over a
given surface.

The proof of associativity is rather lenghty and technical, and it can be
skipped at first reading, since nothing from the following discussion (aside
from the result itself) will be used in later sections. On the other hand,
before we start the proof, let me briefly connect our result to the
literature on deformation quantization \cite{bayen}\cite{lecomte}\cite
{fedosov}\cite{maxim}\cite{guillemin}, where similar products have been
stiudied at length.

The theory of deformation quantization starts with a choice of a real
manifold $\Sigma $, together with a Poisson structure - i.e. an
antisymmetric tensor $\omega ^{ij}$ which satisfies 
\begin{equation}
\omega ^{\alpha \delta }\partial _{\delta }\omega ^{\beta \gamma }+\omega
^{\beta \delta }\partial _{\delta }\omega ^{\gamma \alpha }+\omega ^{\gamma
\delta }\partial _{\delta }\omega ^{\alpha \beta }=0.  \label{eq100}
\end{equation}
(if $\det \omega \neq 0$, then the manifold is symplectic). One also
considers the space $\frak{G}[[\varepsilon ]]$ of formal power series 
\begin{equation*}
A=A_{0}+A_{1}\varepsilon +A_{2}\varepsilon ^{2}+\cdots ,
\end{equation*}
where the $A_{i}$ are functions on the manifold $\Sigma $, and $\varepsilon $
is a quantization parameter. Given two elements $A$ and $B$ in $\frak{G}%
[[\varepsilon ]]$, one may use $\omega ^{ij}$ to define a Poisson bracket 
\begin{equation*}
\{A,B\}=\omega ^{ij}\partial _{i}A\partial _{j}B,
\end{equation*}
which satisfies the Jacobi identity thanks to (\ref{eq100}). The main
problem of deformation quantization in then to define, on the space $\frak{G}%
[[\varepsilon ]]$, an associative product $\star $ 
\begin{equation}
A\star B=\sum_{n=0}^{\infty }\varepsilon ^{n}S_{n}(A,B)  \label{eq105}
\end{equation}
such that

\begin{enumerate}
\item  The $S_{n}$ are bilinear local functionals of $A,B$ ($S_{n}$ depends
only on $A$, $B$, and their derivatives up to a finite order).

\item  $S_{0}(A,B)=AB.$

\item  $S_{1}(A,B)-S_{1}(B,A)=\{A,B\}.$
\end{enumerate}

We can now show that, if we specialize to the case of a symplectic manifold $%
\Sigma $ of real dimension two, then equation (\ref{eq101}) can be used to
determine a class of solutions to the problem of deformation quantization.
We start by choosing, on the manifold $\Sigma $, a complex structure (this
can always be done since the manifold is orientable and has dimension two).
We then define 
\begin{equation*}
P_{1}=\frac{1}{\varepsilon \omega ^{z\overline{z}}}
\end{equation*}
and the higher $P_{n}$ according to the recursion formula (\ref{eq102}). It
is then an easy matter to show that properties $1$-$2$-$3$ above are
satisfied. Let me just conclude this brief discussion on deformation
quantization with a few remarks:

\begin{enumerate}
\item  Using (\ref{eq102}) one can show that 
\begin{equation*}
\varepsilon P_{n}=\frac{n}{\omega ^{z\overline{z}}}+o(\varepsilon ).
\end{equation*}
Therefore the $n$-th term in (\ref{eq101}), although it is of order $%
\varepsilon ^{n}$, also contains terms with higher powers of $\varepsilon $
and therefore does not correspond to the $n$-th term in (\ref{eq105}). One
has to carefully keep track of powers of $\varepsilon $ to go from the
simpler and geometrically clear expression (\ref{eq101}) to the standard
form of deformation quantization (\ref{eq105}). For example, if we denote
for simplicity 
\begin{equation*}
\omega =\omega ^{z\overline{z}}
\end{equation*}
then the first few terms in (\ref{eq105}) read 
\begin{eqnarray}
A\star B &=&AB+\varepsilon \omega \partial A\overline{\partial }B+\frac{%
\varepsilon ^{2}}{2}(\partial \omega \partial A)(\overline{\partial }\omega 
\overline{\partial }B)+  \label{eq110} \\
&&+\frac{\varepsilon ^{3}}{6}\frac{1}{\omega }\left( \partial \omega
\partial \omega \partial A\right) \left( \overline{\partial }\omega 
\overline{\partial }\omega \overline{\partial }B\right) +  \notag \\
&&+\frac{\varepsilon ^{3}}{4}\left( \partial \overline{\partial }\omega
\right) \left( \partial \omega \partial A\right) \left( \overline{\partial }%
\omega \overline{\partial }B\right) -  \notag \\
&&-\frac{\varepsilon ^{3}}{4}\frac{1}{\omega }(\partial \omega )(\overline{%
\partial }\omega )\left( \partial \omega \partial A\right) \left( \overline{%
\partial }\omega \overline{\partial }B\right) +o(\varepsilon ^{4})  \notag
\end{eqnarray}

\item  If one chooses a different complex structure on $\Sigma $ one gets a
different $\star $ product. It is, on the other hand, know \cite{maxim} that
there is a one-to-one correspondence between Poisson structures and
star-products (up to equivalences). Therefore one expects that the various
products related to different complex structures on $\Sigma $ should be
gauge transformations (in the sense of \cite{bayen}) of each-other.

\item  Finally let me note that the expression (\ref{eq101}) cannot be
extended to the case of Poisson manifolds, as can be easily seen from
equation (\ref{eq110}).
\end{enumerate}

Let us now move to the actual proof of associativity of $\star $, and let us
start by choosing three functions $A,B,C$ in $\frak{G}$ and by considering
the product $\left( A\star B\right) \star C$, which we can write as 
\begin{eqnarray*}
\left( A\star B\right) \star C\ &=&\sum_{b,c=0}^{\infty }P_{1}\cdots P_{c}
\left( \frac{1}{P_{c}}\overline{\partial }\cdots \frac{1}{P_{1}}\overline{%
\partial }C\right)\\
&&\left(\frac{1}{P_{c}}\partial \cdots \frac{1}{P_{1}}\partial
\left( \frac{1}{P_{b}}\partial \cdots \frac{1}{P_{1}}\partial A\right)
\left( P_{1}\cdots P_{b}\frac{1}{P_{b}}\overline{\partial }\cdots \frac{1}{%
P_{1}}\overline{\partial }B\right)\right).
\end{eqnarray*}
Let us focus our attention on a specific summand in the above expression for
fixed $b,c$. In particular we wish to analyze in detail the expression 
\begin{equation}
\frac{1}{P_{c}}\partial \cdots \underbrace{\frac{1}{P_{1}}\partial }\left( 
\frac{1}{P_{b}}\partial \cdots \frac{1}{P_{1}}\partial A\right) \left(
P_{1}\cdots P_{b}\frac{1}{P_{b}}\overline{\partial }\cdots \frac{1}{P_{1}}%
\overline{\partial }B\right) .  \label{eq1}
\end{equation}
The first holomorphic derivative $\frac{1}{P_{1}}\partial $ (underlined by $%
\underbrace{\,}$ ) can act on either the first or on the second parenthesis.
In the first case we write the result as 
\begin{equation*}
\frac{1}{P_{c}}\partial \cdots \frac{1}{P_{2}}\partial \left( \frac{1}{%
P_{b+1}}\partial \frac{1}{P_{b}}\partial \cdots \frac{1}{P_{1}}\partial
A\right) \left( P_{2}\cdots P_{b+1}\frac{1}{P_{b}}\overline{\partial }\cdots 
\frac{1}{P_{1}}\overline{\partial }B\right) ,
\end{equation*}
where we have moved the factor of $\frac{1}{P_{1}}$ to the second
parenthesis and we have multiplied and divided by $P_{b+1}$ in order to
maintain the general form of the first parenthesis. In the second case, on
the other hand, we get 
\begin{equation*}
\frac{1}{P_{c}}\partial \cdots \frac{1}{P_{2}}\partial \left( \frac{1}{%
P_{b}}\partial \cdots \frac{1}{P_{1}}\partial A\right)
\left( P_{2}\cdots P_{b+1}\frac{1}{P_{1}\cdots P_{b}}\partial P_{1}\cdots
P_{b}\frac{1}{P_{b}}\overline{\partial }\cdots \frac{1}{P_{1}}\overline{%
\partial }B\right) .
\end{equation*}
The reasons for rewriting the second parenthesis in this way will become
clear later. In a similar way we can distribute the action of the various
holomorphic derivatives on either the left or on the right parenthesis. A
convenient way to summarize the result is as follows. Let $s_{i}\,\
(i=1,\cdots ,c)$, $s_{i}\in \{L,R\}$, be a string indicating whether the $i$%
-th holomorphic derivative $\frac{1}{P_{i}}\partial $ should act on the left
or on the right parenthesis. Also let $L(s)$ be the total number of indices $%
i$ such that $s_{i}=L$. Expression (\ref{eq1}) can then be written as a sum
over all possible choices of $s_{i}$ 
\begin{equation*}
\sum_{\{s_{i}\}_{i=1}^{c}}\left( \frac{1}{P_{b+L(s)}}\partial \cdots \frac{1%
}{P_{1}}\partial A\right) K_{b,c}(s),
\end{equation*}
where the objects $K_{b,c}(s)$ are constructed using the following
algorithm. Start with 
\begin{equation*}
P_{l}\cdots P_{h}\frac{1}{P_{b}}\overline{\partial }\cdots \frac{1}{P_{1}}%
\overline{\partial }B
\end{equation*}
with $l=1$ and $h=b$. If $s_{1}=L$, divide by $P_{l}$ and multiply by $%
P_{h+1}$, and then raise both $l$ and $h$ by one unit. If $s_{1}=R$, then
act on the hole expression with $\partial $, divide by $P_{l}\cdots P_{h}$
and then multiply by $P_{l+1}\cdots P_{h}$. In this second case we just
increase $l$ by one, leaving $h$ fixed. We then repeat this process for $%
s_{2},\cdots ,s_{c}$. The final result will be $K_{b,c}(s)$\footnote{%
We are using the convention that $P_{a}\cdots P_{a}=P_{a}$, $P_{a+1}\cdots
P_{a}=1$, $P_{a+2}\cdots P_{a}=1/P_{a+1}$, etc.}. Let us note that,
regardless of the specific choice of sequence $s_{i}$, at the end of the
algorithm described above, the values of $l$ and $h$ will be respectively $%
1+c$ and $b+L(s)$. Therefore the expression for $K_{b,c}(s)$ will be of the
form 
\begin{equation*}
K_{b,c}(s)=P_{c+1}\cdots P_{b+L(s)}\times \cdots ,
\end{equation*}
where $\cdots $ depends on the specific $s$. We can now rewrite the full
expression for $\left( A\star B\right) \star C$ as 
\begin{equation*}
\sum_{b,c=0}^{\infty }\sum_{\{s_{i}\}_{i=1}^{c}}\left( \frac{1}{P_{b+L(s)}}%
\partial \cdots \frac{1}{P_{1}}\partial A\right) \left( P_{1}\cdots
P_{c}K_{b,c}(s)\right) \left( \frac{1}{P_{c}}\overline{\partial }\cdots 
\frac{1}{P_{1}}\overline{\partial }C\right) .
\end{equation*}
In the above expression we change index by calling $a=b+L(s)$, and we
finally arrive at the following formula 
\begin{equation*}
\left( A\star B\right) \star C=\sum_{a,c=0}^{\infty }\left( \frac{1}{P_{a}}%
\partial \cdots \frac{1}{P_{1}}\partial A\right) T_{a,c}\left( \frac{1}{P_{c}%
}\overline{\partial }\cdots \frac{1}{P_{1}}\overline{\partial }C\right) ,
\end{equation*}
where 
\begin{equation}
T_{a,c}=\sum_{\substack{ \{s_{i}\}_{i=1}^{c}  \\ L(s)\leq a}}P_{1}\cdots
P_{c}K_{a-L(s),c}(s).  \label{eq2}
\end{equation}
Let us consider the expression for $T_{a,c}$. The simplest case is when $c=0$%
. It is immediate to show that 
\begin{equation*}
T_{a,0}=P_{1}\cdots P_{a}\frac{1}{P_{a}}\overline{\partial }\cdots \frac{1}{%
P_{1}}\overline{\partial }B
\end{equation*}
and therefore that 
\begin{equation*}
T_{a,0}=P_{1}\cdots P_{a-1}\overline{\partial }\frac{1}{P_{1}\cdots P_{a-1}}%
T_{a-1,0}.
\end{equation*}
The next simplest case is when $a=0$. Then $s_{i}=R$ for all $i$, and it is
easy to show, following the algorithm described above, that 
\begin{equation*}
T_{0,c}=P_{1}\cdots P_{c}\frac{1}{P_{c}}\partial \cdots \frac{1}{P_{1}}%
\partial B.
\end{equation*}
The corresponding recursion relation reads 
\begin{equation*}
T_{0,c}=P_{1}\cdots P_{c-1}\partial \frac{1}{P_{1}\cdots P_{c-1}}T_{0,c-1}.
\end{equation*}
We now move back to the general case, and start to analyze equation (\ref
{eq2}) for $c,a>0$. We can break the sum in (\ref{eq2}) in two parts and
rewrite equation (\ref{eq2}) as 
\begin{eqnarray}
T_{a,c} &=&\sum_{\substack{ \{s_{i}\}_{i=1}^{c}  \\ L(s)\leq a}}P_{1}\cdots
P_{c}K_{a-L(s),c}(s)=  \label{eq3} \\
&=&\sum_{\substack{ \{s_{i}\}_{i=1}^{c}  \\ s_{c}=L,\,L(s)\leq a}}%
P_{1}\cdots P_{c}K_{a-L(s),c}(s)+\sum_{\substack{ \{s_{i}\}_{i=1}^{c}  \\ %
s_{c}=R,\,L(s)\leq a}}P_{1}\cdots P_{c}K_{a-L(s),c}(s)  \notag
\end{eqnarray}
If one follows carefully the algorithm defining the functions $K_{b,c}$'s,
one can show that the first term in the above expression can be rewritten as 
\begin{eqnarray*}
&&\sum_{\substack{ \{s_{i}\}_{i=1}^{c}  \\ s_{c}=L,\,L(s)\leq a}}P_{1}\cdots
P_{c}\frac{P_{a}}{P_{c}}K_{a-L(s),c-1}(s) = \\
&&=\sum_{\substack{ \{s_{i}\}_{i=1}^{c-1}  \\ L(s)\leq a-1}}P_{a}P_{1}\cdots
P_{c-1}K_{a-L(s)-1,c-1}(s) = P_{a}T_{a-1,c-1}.
\end{eqnarray*}
The second term of (\ref{eq3}) can, on the other hand be rewritten as
\begin{eqnarray*}
&& \sum_{\substack{ \{s_{i}\}_{i=1}^{c}  \\ s_{c}=R,\,L(s)\leq a}}P_{1}\cdots
P_{c}\frac{1}{P_{c}}\partial K_{a-L(s),c-1}(s) = \\
&& = \sum_{\substack{ \{s_{i}\}_{i=1}^{c-1}  \\ \,L(s)\leq a}}P_{1}\cdots
P_{c-1}\partial \frac{1}{P_{1}\cdots P_{c-1}}P_{1}\cdots
P_{c-1}K_{a-L(s),c-1}(s) = \\
&& = P_{1}\cdots P_{c-1}\partial \frac{1}{P_{1}\cdots P_{c-1}}T_{a,c-1}.
\end{eqnarray*}
Combining the above two terms, we finally arrive at the general recursion
relation, valid for$\ c,a>0$, 
\begin{equation*}
T_{a,c}=P_{a}T_{a-1,c-1}+P_{1}\cdots P_{c-1}\partial \frac{1}{P_{1}\cdots
P_{c-1}}T_{a,c-1}.
\end{equation*}
Let us summarize what we have found up to now in the following

\begin{claim}
\label{claim3}We can rewrite the expression for $\left( A\star B\right)
\star C$ as 
\begin{equation*}
\left( A\star B\right) \star C=\sum_{a,c=0}^{\infty }\left( \frac{1}{P_{a}}%
\partial \cdots \frac{1}{P_{1}}\partial A\right) T_{a,c}\left( \frac{1}{P_{c}%
}\overline{\partial }\cdots \frac{1}{P_{1}}\overline{\partial }C\right) ,
\end{equation*}
where the functions $T_{a,c}$ satisfy the following recursion relations 
\begin{equation*}
\begin{array}[b]{ll}
T_{0,0}=B &  \\ 
T_{0,c}=P_{1}\cdots P_{c-1}\partial \frac{1}{P_{1}\cdots P_{c-1}}T_{0,c-1} & 
(c>0) \\ 
T_{a,0}=P_{1}\cdots P_{a-1}\overline{\partial }\frac{1}{P_{1}\cdots P_{a-1}}%
T_{a-1,0} & (a>0) \\ 
T_{a,c}=P_{a}T_{a-1,c-1}+P_{1}\cdots P_{c-1}\partial \frac{1}{P_{1}\cdots
P_{c-1}}T_{a,c-1} & (a,c>0)
\end{array}
\end{equation*}
\end{claim}

In a completely symmetric way we can also prove the following

\begin{claim}
\label{claim4}We can rewrite the expression for $A\star (B\star C)$ as 
\begin{equation*}
A\star (B\star C)=\sum_{a,c=0}^{\infty }\left( \frac{1}{P_{a}}\partial
\cdots \frac{1}{P_{1}}\partial A\right) \widetilde{T}_{a,c}\left( \frac{1}{%
P_{c}}\overline{\partial }\cdots \frac{1}{P_{1}}\overline{\partial }C\right)
,
\end{equation*}
where the functions $\widetilde{T}_{a,c}$ satisfy the following recursion
relations 
\begin{equation*}
\begin{array}{ll}
\widetilde{T}_{0,0}=B &  \\ 
\widetilde{T}_{0,c}=P_{1}\cdots P_{c-1}\partial \frac{1}{P_{1}\cdots P_{c-1}}%
\widetilde{T}_{0,c-1} & (c>0) \\ 
\widetilde{T}_{a,0}=P_{1}\cdots P_{a-1}\overline{\partial }\frac{1}{%
P_{1}\cdots P_{a-1}}\widetilde{T}_{a-1,0} & (a>0) \\ 
\widetilde{T}_{a,c}=P_{c}\widetilde{T}_{a-1,c-1}+P_{1}\cdots P_{a-1}%
\overline{\partial }\frac{1}{P_{1}\cdots P_{a-1}}\widetilde{T}_{a-1,c} & 
(a,c>0)
\end{array}
\end{equation*}
\end{claim}

It is then clear that, to complete the proof of associativity of the product 
$\star $, it suffices to show that 
\begin{equation*}
T_{a,c}=\widetilde{T}_{a,c}
\end{equation*}
for all $a,c\geq 0$. We break the proof in five steps.
\vskip.3cm
\textit{Step 1.} We first note that the recursion relations in Claims \ref
{claim3} and \ref{claim4} immediately imply that 
\begin{equation*}
T_{a,0}=\widetilde{T}_{a,0}\,\ \ \ \ \ \ \ \ \ T_{0,c}=\widetilde{T}_{0,c}
\end{equation*}
for any $a,c\geq 0$.
\vskip.3cm
\textit{Step 2. }We then prove that 
\begin{equation*}
T_{1,1}=\widetilde{T}_{1,1}.
\end{equation*}
This can be shown by simply noting that 
\begin{equation*}
T_{1,1}=P_{1}B+\partial T_{1,0}=P_{1}B+\partial \overline{\partial }B
\end{equation*}
and that 
\begin{equation*}
\widetilde{T}_{1,1}=P_{1}B+\overline{\partial }T_{0,1}=P_{1}B+\overline{%
\partial }\partial B.
\end{equation*}
\vskip.3cm
\textit{Step 3. }We prove now that 
\begin{equation*}
T_{a,1}=\widetilde{T}_{a,1}
\end{equation*}
for $a\geq 2$, by induction on $a$. Using the recursion relations in Claim 
\ref{claim3} we can write 
\begin{eqnarray*}
T_{a,1} &=&P_{a}T_{a-1,0}+\partial T_{a,0}= \\
&=&P_{a}T_{a-1,0}+\partial P_{1}\cdots P_{a-1}\overline{\partial }\frac{1}{%
P_{1}\cdots P_{a-1}}T_{a-1,0}
\end{eqnarray*}
On the other hand the equivalent recursion relation for $\widetilde{T}_{a,1}$
can be combined with the induction hypothesis $T_{a-1,1}=\widetilde{T}%
_{a-1,1}$and the fact that $T_{a-1,0}=\widetilde{T}_{a-1,0}$ to show that 
\begin{equation*}
\widetilde{T}_{a,1}=P_{1}T_{a-1,0}+P_{1}\cdots P_{a-1}\overline{\partial }%
\frac{1}{P_{1}\cdots P_{a-1}}T_{a-1,1}.
\end{equation*}
In the above expression we then use the recursion relation for $T_{a-1,1}$
and we get 
\begin{eqnarray*}
\widetilde{T}_{a,1} &=&P_{1}T_{a-1,0}+P_{1}\cdots P_{a-1}\overline{\partial }%
\frac{1}{P_{1}\cdots P_{a-1}}(P_{a-1}T_{a-2,0}+\partial T_{a-1,0})= \\
&=&P_{1}T_{a-1,0}+P_{a-1}P_{1}\cdots P_{a-2}\overline{\partial }\frac{1}{%
P_{1}\cdots P_{a-2}}T_{a-2,0}+ \\
&&+P_{1}\cdots P_{a-1}\overline{\partial }\frac{1}{P_{1}\cdots P_{a-1}}%
\partial T_{a-1,0} \\
&=&(P_{1}+P_{a-1})T_{a-1,0}+P_{1}\cdots P_{a-1}\overline{\partial }\frac{1}{%
P_{1}\cdots P_{a-1}}\partial T_{a-1,0}
\end{eqnarray*}
We are now almost done. In order to show that $\widetilde{T}_{a,1}=T_{a,1}$,
we first note that we can use the properties of the tensors $P_{n}$ to
simplify the following difference 
\begin{eqnarray*}
&&\partial P_{1}\cdots P_{a-1}\overline{\partial }\frac{1}{P_{1}\cdots
P_{a-1}}T_{a-1,0}-P_{1}\cdots P_{a-1}\overline{\partial }\frac{1}{%
P_{1}\cdots P_{a-1}}\partial T_{a-1,0} \\
&=&\left[ \partial ,P_{1}\cdots P_{a-1}\left[ \overline{\partial },\frac{1}{%
P_{1}\cdots P_{a-1}}\right] \right] T_{a-1,0}=-\left( \partial \overline{%
\partial }\ln P_{1}\cdots P_{a-1}\right) T_{a-1,0} \\
&=&(P_{1}+P_{a-1}-P_{a})T_{a-1,0}.
\end{eqnarray*}
But this shows that $\widetilde{T}_{a,1}-T_{a,1}=0$, thus concluding the
inductive step.
\vskip.3cm
\textit{Step 4.} In a way completely equivalent to \textit{Step 3} we can
also prove that 
\begin{equation*}
T_{1,c}=\widetilde{T}_{1,c}
\end{equation*}
for $c\geq 2$.
\vskip.3cm
\textit{Step 5.} Finally we will show that 
\begin{equation*}
T_{a,c}=\widetilde{T}_{a,c}
\end{equation*}
for $a,c\geq 2$. The proof will be again by induction on both $a$ and $c$,
and will be very similar to \textit{Step 3 }and 4, even though it will be
considerably more complex notationally.

We start by writing the recursion relation satisfied by $T_{a,c}$%
\begin{equation*}
T_{a,c}=P_{a}T_{a-1,c-1}+P_{1}\cdots P_{c-1}\partial \frac{1}{P_{1}\cdots
P_{c-1}}T_{a,c-1}.
\end{equation*}
In the above expression, the induction hypothesis allows us to go from $T$'s
to $\tilde{T}$'s, and we can then use the recursion relations for $%
\widetilde{T}$ to write 
\begin{eqnarray*}
T_{a,c} &=&P_{a}P_{c-1}T_{a-2,c-2}+P_{a}P_{1}\cdots P_{a-2}\overline{%
\partial }\frac{1}{P_{1}\cdots P_{a-2}}T_{a-2,c-1}+ \\
&&+P_{c-1}P_{1}\cdots P_{c-2}\partial \frac{1}{P_{1}\cdots P_{c-2}}%
T_{a-1,c-2}+ \\
&&+P_{1}\cdots P_{c-1}\partial \frac{P_{1}\cdots P_{a-1}}{P_{1}\cdots P_{c-1}%
}\overline{\partial }\frac{1}{P_{1}\cdots P_{a-1}}T_{a-1,c-1}
\end{eqnarray*}
On the other hand one could start with 
\begin{equation*}
\widetilde{T}_{a,c}=P_{c}\widetilde{T}_{a-1,c-1}+P_{1}\cdots P_{a-1}%
\overline{\partial }\frac{1}{P_{1}\cdots P_{a-1}}\widetilde{T}_{a-1,c}
\end{equation*}
and use the induction hypothesis and the recursion formula to write 
\begin{eqnarray*}
\widetilde{T}_{a,c} &=&P_{c}P_{a-1}T_{a-2,c-2}+P_{c}P_{1}\cdots
P_{c-2}\partial \frac{1}{P_{1}\cdots P_{c-2}}T_{a-1,c-2}+ \\
&&+P_{a-1}P_{1}\cdots P_{a-2}\overline{\partial }\frac{1}{P_{1}\cdots P_{a-2}%
}T_{a-2,c-1}+ \\
&&+P_{1}\cdots P_{a-1}\overline{\partial }\frac{P_{1}\cdots P_{c-1}}{%
P_{1}\cdots P_{a-1}}\partial \frac{1}{P_{1}\cdots P_{c-1}}T_{a-1,c-1}
\end{eqnarray*}
At this point we just need to consider the difference $\widetilde{T}%
_{a,c}-T_{a,c}$ and to show that it vanishes. The complete expression for $%
\widetilde{T}_{a,c}-T_{a,c}$ can be written as the sum of two parts, each of
which can be simplified using the properties of the tensors $P_{n}$. On one
hand\ we can consider the difference 
\begin{eqnarray}
&&P_{1}\cdots P_{a-1}\overline{\partial }\frac{P_{1}\cdots P_{c-1}}{%
P_{1}\cdots P_{a-1}}\partial \frac{1}{P_{1}\cdots P_{c-1}}T_{a-1,c-1}-
\label{eq4} \\
&&-P_{1}\cdots P_{c-1}\partial \frac{P_{1}\cdots P_{a-1}}{P_{1}\cdots P_{c-1}%
}\overline{\partial }\frac{1}{P_{1}\cdots P_{a-1}}T_{a-1,c-1}  \notag \\
&=&-P_{1}\cdots P_{c-1}\left[ \partial ,\frac{P_{1}\cdots P_{a-1}}{%
P_{1}\cdots P_{c-1}}\left[ \overline{\partial },\frac{P_{1}\cdots P_{c-1}}{%
P_{1}\cdots P_{a-1}}\right] \right] \frac{1}{P_{1}\cdots P_{c-1}}T_{a-1,c-1}=
\notag \\
&=&\partial \overline{\partial }\ln \left( \frac{P_{1}\cdots P_{a-1}}{%
P_{1}\cdots P_{c-1}}\right) T_{a-1,c-1}=\left(
P_{a}-P_{a-1}-P_{c}+P_{c-1}\right) T_{a-1,c-1}  \notag
\end{eqnarray}
On the other hand, we compute the difference 
\begin{eqnarray*}
&&P_{c}P_{a-1}T_{a-2,c-2}+P_{c}P_{1}\cdots P_{c-2}\partial \frac{1}{%
P_{1}\cdots P_{c-2}}T_{a-1,c-2}+ \\
&&+P_{a-1}P_{1}\cdots P_{a-2}\overline{\partial }\frac{1}{P_{1}\cdots P_{a-2}%
}T_{a-2,c-1}- \\
&&-P_{a}P_{c-1}T_{a-2,c-2}-P_{a}P_{1}\cdots P_{a-2}\overline{\partial }\frac{%
1}{P_{1}\cdots P_{a-2}}T_{a-2,c-1}- \\
&&-P_{c-1}P_{1}\cdots P_{c-2}\partial \frac{1}{P_{1}\cdots P_{c-2}}%
T_{a-1,c-2}\\
&=&(P_{c}-P_{c-1})P_{1}\cdots P_{c-2}\partial \frac{1}{P_{1}\cdots P_{c-2}}%
T_{a-1,c-2}+ \\
&&+(P_{a-1}-P_{a})P_{1}\cdots P_{a-2}\overline{\partial }\frac{1}{%
P_{1}\cdots P_{a-2}}T_{a-2,c-1}+ \\
&&+(P_{c}P_{a-1}-P_{c-1}P_{a-1}+P_{c-1}P_{a-1}-P_{a}P_{c-1})T_{a-2,c-2} \\
&=&\left( P_{c}-P_{c-1}-P_{a}+P_{a-1}\right) T_{a-1,c-1}
\end{eqnarray*}
which exactly cancels expression (\ref{eq4}), thus concluding the proof of
the following

\begin{theorem}
For all $a,b\geq 0$%
\begin{equation*}
\widetilde{T}_{a,c}=T_{a,c}.
\end{equation*}
Therefore 
\begin{equation*}
\left( A\star B\right) \star C=A\star \left( B\star C\right) .
\end{equation*}
\end{theorem}

\section{\label{sec4}\label{sb}From Star Products to Bergman Projections}

In section \ref{bs} we started with the choice of a specific holomorphic
line bundle $S$ on $\Sigma $ together with a measure $C$, and we constructed
a star-product $\star $, which is defined only in terms of the curvature
tensor $R$. In this section we will analyze the inverse problem of
reconstructing $S$ and $C$ given a known $R$.

Let then $R$ be given on the surface. Using Dolbeault's lemma we can find a
cover $\mathcal{U}_{i}$ of $\Sigma $ and real functions $\mathcal{L}_{i}$ on 
$\mathcal{U}_{i}$ such that, on the $i$-th patch, $R=\partial \overline{%
\partial }\mathcal{L}_{i}$. On the intersections $\mathcal{U}_{i}\cap 
\mathcal{U}_{j}$ we have $\mathcal{L}_{j}=\mathcal{L}_{i}+\lambda _{ij}+%
\overline{\lambda }_{ij}$, for some holomorphic functions $\lambda _{ij}$,
which are defined up to an imaginary constant. On the triple intersections $%
\mathcal{U}_{i}\cap \mathcal{U}_{j}\cap \mathcal{U}_{k}$, we then have that $%
\lambda _{ij}+\lambda _{jk}=\lambda _{ik}+2\pi i\,n_{ijk}$, where the $n$'s
are constant real numbers. Let us suppose, for now, that we can redefine the
functions $\lambda _{ij}$ in such a way that the numbers $n_{ijk}$ are
actually integers. In this case then, the functions $g_{ij}=e^{\lambda
_{ij}} $ define a holomorphic line bundle\footnote{%
We are not concerned here with questions of uniqueness.}, which we will
denote by $T$. If we recall that $R=\partial \overline{\partial }\ln C$, it
is natural to let $C=e^{\mathcal{L}}$, which is a section of $T\otimes 
\overline{T}$, and to conclude the inverse construction by letting $%
S=T\otimes K^{-1}$.

Let us compute the degree of the line bundle $T$. Recall that $d-\Gamma
\,dz=d-\partial \ln C\,dz$ defines a covariant derivative for sections of $T$%
. The first Chern class of the line bundle is then given in terms of the
curvature $2$-form $i\overline{\partial }\Gamma \,dz\wedge d\overline{z}$ by 
\begin{equation*}
c_{1}(T)=-\frac{1}{2\pi i}R\,dz\wedge d\overline{z},
\end{equation*}
and therefore we can compute the degree of $T$ as 
\begin{equation}
\deg (T)=\int_{\Sigma }c_{1}(T)=-\frac{1}{2\pi i}\int_{\Sigma }R\,dz\wedge d%
\overline{z}.  \label{eq5}
\end{equation}
Given a generic $R$, the above integral will not give an integer, and
therefore the inversion problem cannot be solved. On the other hand, it is a
known fact that, if the integral (\ref{eq5}) does yield a number in $\mathbb{%
Z}$, then we can redefine the $\lambda $'s introduced above (not necessarily
uniquely) so that the numbers $n_{ijk}$ are all integers, and the problem of
finding $C$ and $S$ can be solved constructively, as I have previously shown.

\section{\label{sec5}\label{k}Holomorphic Curves on Compact K\"{a}hler Spaces%
}

We will now start to analyze the problem of the quantization of holomorphic
curves embedded in compact spaces. This section, in particular, is devoted
to the general case, where the target space is a generic K\"{a}hler
manifold. The discussion will serve as an introduction to the next section,
where we restrict our attention to curves embedded in complex tori, and
where we relate our quantization procedure to Yang-Mills configurations on
the dual tori.

Let us start with a generic Riemann surface $\Sigma $ of genus $g$, embedded
holomorphically in a compact K\"{a}hler manifold $M$. The inclusion map $%
\rho :\Sigma \rightarrow M$ induces on the surface $\Sigma $ a K\"{a}hler
form, which we will denote by 
\begin{equation*}
\mu =\frac{i}{2}Q\,dz\wedge d\overline{z}
\end{equation*}
The integral of $\,\mu $ over the surface 
\begin{equation}
A=\int_{\Sigma }\mu  \label{eq300}
\end{equation}
is nothing but the area of $\Sigma $ considered as a submanifold of the
Riemannian manifold $M$.

We now recall the main result of \cite{aaa}, which can be easily extended to
the present setting. As a function of the quantization parameter $%
\varepsilon $, the curvature $R$ of the associated star-product has an
asymptotic expansion of the form 
\begin{equation}
R(\varepsilon )=-\frac{1}{\varepsilon }Q+\frac{1}{2}\partial \overline{%
\partial }\ln Q+\partial \overline{\partial }\mathcal{G}(\varepsilon ),
\label{eq6}
\end{equation}
where $\mathcal{G}(\varepsilon )$ is a function on $\Sigma $ expressed as a
power series in $\varepsilon $, whose first few terms are 
\begin{equation*}
\mathcal{G}(\varepsilon )=-\frac{\varepsilon }{6}\frac{1}{Q}\partial 
\overline{\partial }\ln Q+\frac{\varepsilon ^{2}}{24}\frac{1}{Q}\partial 
\overline{\partial }\frac{1}{Q}\partial \overline{\partial }\ln Q+\cdots
\end{equation*}
We are now faced with the problem, analyzed in the last section, of finding
the correct quantization $S,C$ as a function of $\varepsilon $. We expect
that not all values of $\varepsilon $ will be allowed, since we must impose
that $\int R(\varepsilon )\,dz\wedge d\overline{z}\in 2\pi i\mathbb{Z}$. On
the other hand we note that, in the expansion (\ref{eq6}), all the terms
with positive powers of $\varepsilon $ are total derivatives and integrate
to zero. Therefore we are left with the simpler quantization condition for $%
\varepsilon $%
\begin{equation*}
\frac{1}{2\pi i}\int_{\Sigma }\left( -\frac{1}{\varepsilon }Q+\frac{1}{2}%
\partial \overline{\partial }\ln Q\right) dz\wedge d\overline{z}\in \mathbb{Z%
}
\end{equation*}
We note that the second term in the integrand is proportional to the
Riemannian curvature on the surface, and the integral can be computed using
the formula of Gauss and Bonnet. The result, $1-g$, is integral and
independent of the embedding. Therefore we finally arrive at the following
quantization condition for $\varepsilon $%
\begin{eqnarray*}
n &=&\frac{1}{\pi \varepsilon }\int_{\Sigma }\mu \\
\varepsilon &=&\frac{1}{\pi }\frac{A}{n}.
\end{eqnarray*}
To find the correct holomorphic line bundle $S$, we start by noticing that
the $2$-form $\mu /A$ can be associated, using the same reasoning as in
section \ref{sb}, to a holomorphic line bundle $L$ of degree 
\begin{equation*}
\deg (L)=1.
\end{equation*}
More precisely, we can find a covering $\mathcal{U}_{i}$ of $\Sigma $ and
real functions $\mathcal{K}_{i}$ on $\mathcal{U}_{i}$ such that $Q=\partial 
\overline{\partial }\mathcal{K}_{i}$ and such that $\mathcal{K}_{j}=\mathcal{%
K}_{i}+\kappa _{ij}+\overline{\kappa }_{ij}$. The line bundle $L$ will then
be defined by the transition functions $f_{ij}=e^{\frac{\pi }{A}\kappa
_{ij}} $. Notice that we can rewrite (\ref{eq6}) in terms of $\mathcal{K}$
as 
\begin{eqnarray*}
R(\varepsilon ) &=&\partial \overline{\partial }\mathcal{L}_{i} \\
\mathcal{L}_{i} &=&-\frac{1}{\varepsilon }\mathcal{K}_{i}+\frac{1}{2}\ln Q+%
\mathcal{G}.
\end{eqnarray*}
Therefore 
\begin{equation*}
C=e^{\mathcal{L}}=e^{-n\frac{\pi }{A}\mathcal{K}}\,\sqrt{Q}\,e^{\mathcal{G}}
\end{equation*}
transforms as a section of $T\otimes \overline{T}$, where 
\begin{equation*}
T=L^{-n}\otimes K^{1/2}
\end{equation*}
and $K^{1/2}$, the square root of the line bundle, is a choice of spin
structure on the surface $\Sigma $. Finally we have that 
\begin{equation*}
S=L^{n}\otimes K^{1/2}.
\end{equation*}
The degree of $S$ can be simply computed to be 
\begin{equation*}
\deg (S)=n\deg (L)+\frac{1}{2}\deg (K)=n+g-1.
\end{equation*}
Finally we wish to compute the dimension 
\begin{equation*}
N(n)=\dim \mathcal{H}
\end{equation*}
of the space $\mathcal{H}$ of holomorphic sections of $S$. This can be done
for large $n$ using the Riemann-Roch theorem. In fact, if $\deg (S)>\deg (K)$%
, or if $n>g-1$, then $h_{1}(S)=0$, and we may write 
\begin{eqnarray*}
\dim \mathcal{H} &=&h_{0}(S)=h_{0}(S)-h_{1}(S)=1-g+\deg (S) \\
N(n) &=&n \\
(n &>&g-1)
\end{eqnarray*}

\section{\label{sec6}\label{tr}\label{sec}Traces of Operators in the $%
\protect\varepsilon \rightarrow 0$ Limit}

In the previous section we have shown how to pass from functions $A$ on $%
\Sigma $ to operators $\mathcal{A}=\mathcal{Q}(A)$ on $\mathcal{H}$ using
Bergman quantization, so that products of operators are associated, at least
asymptotically in $\varepsilon $, to star-products of functions. We now wish
to use the differential representation of the Bergman projection to compute
tr$(\mathcal{A})$ in terms of the original function $A$.

Let $h_{i}$ be an orthonormal basis of $\mathcal{H}$. We compute the trace
of $\mathcal{A}$ as 
\begin{equation*}
\text{tr}(\mathcal{A})=\sum_{i}\left\langle i|\mathcal{A}|i\right\rangle
=\sum_{i}\int_{\Sigma }\overline{h}_{i}(z)h_{i}(z)A(z)\,\Omega (z).
\end{equation*}
Recalling the expression for the Bergman kernel $K(z,w)=\sum_{i}h_{i}(z)%
\overline{h}_{i}(w)$, we may rewrite 
\begin{eqnarray}
\text{tr}(\mathcal{A}) &=&\int_{\Sigma }K(z,z)A(z)\,\Omega (z)\,
\label{eq31} \\
&=&\int_{\Sigma }\mu (z)A(z)\int_{\Sigma }K(z,w)\delta _{z}(w)\Omega (w), 
\notag
\end{eqnarray}
where $\delta _{z}\in \mathcal{V}$ is the distribution with support at the
point $z$ defined by $F(z)=\int_{\Sigma }\mu (w)\delta _{z}(w)F(w)$, where $%
F\in \mathcal{V}$. Note that the second integral in expression (\ref{eq31})
is nothing but the integral representation of the Bergman projection of $%
\delta _{z}$. Therefore we conclude that

\begin{equation*}
\text{tr}(\mathcal{A})=\int_{\Sigma }\mu (z)A(z)\pi (\delta _{z})(z)
\end{equation*}
Let us consider the above expression in the $\varepsilon \rightarrow 0$
limit. We wish to use the differential representation for the projection $%
\pi $, but in order to do so we must regularize the delta function
distribution. Only at the end of the computation we can remove the
regulator. To be concrete we will work in a particular coordinate system $s$
centered at the point $z$ and we regularize the distribution $\delta _{z}$
with a gaussian 
\begin{equation*}
\delta _{z}(s)\rightarrow \delta _{z,\lambda }(s)=\frac{1}{Q(0)}\frac{1}{\pi
\lambda }e^{-s\overline{s}/\lambda }.
\end{equation*}
The projection of $\delta _{z,\lambda }$ is explicitly given by 
\begin{equation}
\pi (\delta _{z,\lambda })(s)=\sum_{n=0}^{\infty }(-1)^{n}\nabla \frac{1}{%
P_{1}}\cdots \nabla \frac{1}{P_{n}}P_{1}\cdots P_{n}\frac{1}{P_{n}}\overline{%
\nabla }\cdots \frac{1}{P_{1}}\overline{\nabla }\delta _{z,\lambda }(s),
\label{eq32}
\end{equation}
where $\nabla =\partial +\Gamma $ and $\overline{\nabla }=\overline{\partial 
}$. Recall that we wish to consider the above expression evaluated at the
point $z$ (at the coordinate $s=0$), in the limit $\lambda \rightarrow 0$.
To extract the most singular part of the contribution we must then act with
the antiholomorphic derivatives $\overline{\partial }$ on the gaussian
function, since every derivative contributes an inverse power of $\lambda $.
In doing so we are also left with a factor of $(-s)^{n}$. Therefore, in
order not to get a vanishing result in the $s\rightarrow 0$ limit, we must
replace $\nabla $ with $\partial $, and we must act with the holomorphic
derivatives on the factor $(-s)^{n}$. Using the fact that $\partial
^{n}(-s)^{n}=(-1)^{n}n!$ and the fact that, to lowest order, $%
P_{n}=-nQ/\varepsilon $, we can then write 
\begin{equation*}
\pi (\delta _{z,\lambda })(0)\backsimeq \frac{1}{\pi \lambda Q}%
\sum_{n=0}^{\infty }(-1)^{n}\left( \frac{\varepsilon }{\lambda }\right) ^{n}%
\frac{1}{Q^{n}}=\frac{1}{\pi \lambda Q+\pi \varepsilon }.
\end{equation*}
In the limit $\lambda \rightarrow 0$ we finally get 
\begin{equation*}
\pi (\delta _{z})\backsimeq \frac{1}{\pi \varepsilon }.
\end{equation*}
We have then concluded that, in the classical limit $\varepsilon \rightarrow
0$, we may compute traces using the formula 
\begin{equation}
\text{tr}(\mathcal{A})\backsimeq \frac{1}{\pi \varepsilon }\int_{\Sigma }\mu
A.\text{ \ \ \ \ \ \ \ \ \ }(\varepsilon \rightarrow 0\text{ limit})
\label{eq33}
\end{equation}
One expects the general expression for the trace to be an asymptotic
expansion in higher powers of $\varepsilon $%
\begin{eqnarray}
\text{tr}(\mathcal{A}) &=&\frac{1}{\pi }\int_{\Sigma }\omega (\varepsilon )A
\label{eq34} \\
\omega (\varepsilon ) &=&\frac{1}{\varepsilon }\mu +\mu _{o}+\varepsilon \mu
_{1}+\cdots ,  \notag
\end{eqnarray}
where the 2-forms $\mu _{i}$ can be computed by looking at subleading terms
in the expansion (\ref{eq32}). One property of the forms $\mu _{i}$ can, on
the other hand, be deduced with very little work. Consider the function $1$
and the corresponding operator $Id_{\mathcal{H}}$. Using the results of
section \ref{k} we may write, for $n>g-1$, 
\begin{equation*}
\text{tr}(Id_{\mathcal{H}})=\dim (\mathcal{H})=n=\frac{A}{\pi \varepsilon }=%
\frac{1}{\pi \varepsilon }\int_{\Sigma }\mu .
\end{equation*}
Therefore the expression (\ref{eq33}) is exact in the case of the function $1%
\footnote{%
Recall that (\ref{eq34}) is an asymptotic expansion in $\varepsilon $, and
therefore it is not sensitive to the fact that $N(n)\neq n$ for $n\leq g-1$.}
$, and we then deduce that 
\begin{equation*}
\int_{\Sigma }\mu _{i}=0\text{ \ \ \ \ \ \ \ \ }(i\geq 0)
\end{equation*}
and therefore that the forms $\mu _{i}$ are all total derivatives.

\section{\label{t}Holomorphic Curves on $T_{4}$}

In this final part of the paper we are going to use the results obtained in
the previous sections to tackle the problem of quantization of holomorphic
curves embedded in complex tori. In particular, for simplicity of notation,
we are going to consider tori of real dimension $4$, even though higher
dimensional examples can be treated with exactly the same techniques.

Let us first fix some notation. Euclidean $4$-space $\mathbb{R}^{4}$, with
the standard flat metric, will be parametrized by coordinates 
\begin{equation*}
x_{a}\text{ \ \ \ \ \ \ }(a=1,\cdots ,4)
\end{equation*}
and will be considered as a complex K\"{a}hler manifold $\mathbb{C}^{2}$,
with a complex structure compatible with the metric. We will be definite and
choose analytic coordinates $z_{i}\,(i=1,2)$ 
\begin{eqnarray*}
z_{1} &=&x_{1}+ix_{2} \\
z_{2} &=&x_{3}+ix_{4}
\end{eqnarray*}
(the various possible choices are parametrized by $SO(4)/U(1)\times U(1)$),
in terms of which the K\"{a}hler form will be 
\begin{eqnarray*}
\mu &=&\frac{i}{2}\partial z_{1}\wedge \overline{\partial }\overline{z}_{1}+%
\frac{i}{2}\partial z_{2}\wedge \overline{\partial }\overline{z}_{2}= \\
&=&dx_{1}\wedge dx_{2}+dx_{3}\wedge dx_{4}.
\end{eqnarray*}
We fix, in $\mathbb{R}^{4}$, a lattice $\Gamma $ of maximal rank generated
by the basis vectors 
\begin{equation*}
a^{A}\text{ \ \ \ \ \ \ }(A=1,\cdots ,4)
\end{equation*}
and we denote by 
\begin{equation*}
T_{4}=\mathbb{R}^{4}/\Gamma
\end{equation*}
the quotient $4$-torus, which inherits from $\mathbb{C}^{2}$ the K\"{a}hler
structure. In what follows, we will also need to consider the torus 
\begin{equation*}
\tilde{T}_{4}=\mathbb{R}^{4}/\tilde{\Gamma}
\end{equation*}
dual to $T_{4}$. In the above expression $\tilde{\Gamma}$ denotes the
lattice dual to $\Gamma $, which is generated by the vectors 
\begin{equation*}
b_{B}\text{ \ \ \ \ \ \ }(B=1,\cdots ,4)
\end{equation*}
satisfying 
\begin{equation*}
a^{A}\cdot b_{B}=\delta _{B}^{A}.
\end{equation*}

Let us now consider a curve $\Sigma $ embedded holomorphically in $T_{4}$.
We are clearly within the general framework described in section \ref{k},
with $M=T_{4}$. With a slight abuse of notation we are denoting with $\mu $
both the K\"{a}hler form in the target space and the K\"{a}hler form induced
on the surface $\Sigma $. We will also loosely talk about coordinate
functions $X:\Sigma \rightarrow \mathbb{R}^{4}$, but we will have to keep in
mind that they are multivalued functions, defined only up to elements of the
lattice $\Gamma $%
\begin{equation*}
X\sim X+n_{A}a^{A}.\text{\ \ \ \ \ }(n_{A}\in \mathbb{Z})
\end{equation*}
The differentials $dX_{a}$ are, on the other hand, well defined on the
surface $\Sigma $, and one can write the K\"{a}hler form on $\Sigma $ as 
\begin{equation*}
\mu =dX_{1}\wedge dX_{2}+dX_{3}\wedge dX_{4}.
\end{equation*}
As we described in detail in section \ref{k}, we have on the surface $\Sigma 
$, as a function of the quantization parameter $\varepsilon $, a well
defined star product. In particular we notice that, although the product $%
X_{a}\star X_{b}$ is ill defined, the commutator $\left[ X_{a},X_{b}\right]
=X_{a}\star X_{b}-X_{b}\star X_{a}$ only depends on derivatives of the
coordinate functions, and therefore represents, asymptotically in $%
\varepsilon $, a function on the surface $\Sigma $. Moreover we recall from 
\cite{aaa} that, if we define $Z_{1}=X_{1}+iX_{2}$, $Z_{2}=X_{3}+iX_{4}$,
the deformation (\ref{eq6}) of the curvature $R(\varepsilon )$ was chosen so
that $\left[ Z_{1},Z_{2}\right] =0$ and that $[Z_{1},\overline{Z}%
_{1}]+[Z_{2},\overline{Z}_{2}]=-\varepsilon $. Rewriting these relations in
terms of the euclidean coordinates we have that 
\begin{eqnarray}
-2\pi i\left( \left[ X_{1},X_{2}\right] +\left[ X_{3},X_{4}\right] \right)
&=&-\pi \varepsilon  \label{eq42} \\
\left[ X_{1},X_{3}\right] -\left[ X_{2},X_{4}\right] &=& \left[ X_{1},X_{4}\right] +\left[ X_{2},X_{3}\right] = 0.  \notag
\end{eqnarray}
The functions $X_{a}$ cannot be directly quantized, since they are
multivalued on the surface $\Sigma $. To avoid this problem let us proceed
formally and consider the objects 
\begin{equation*}
U(y)=e^{2\pi i\,y\cdot X},
\end{equation*}
where $y$ are coordinates in $\mathbb{R}^{4}$. In the above expression, and
in the ones that follow, all the products between functions should be
considered as star-products. In particular 
\begin{equation*}
e^{A}=1+A+\frac{1}{2}A\star A+\frac{1}{6}A\star A\star A+\cdots .
\end{equation*}
For a generic $y$ the objects $U(y)$ do not represent well-defined functions
on $\Sigma $. On the other hand, let us take the following point of view.
Let us call $\frak{g}$ the algebra (with respect to the product $\star $) of
real functions on $\Sigma $ ($\frak{G=g}_{\mathbb{C}}$) and let us consider
it as the Lie algebra of a group $G$ (in the $\varepsilon \rightarrow 0$
limit, $G$ becomes the group of $\mu $-area preserving diffeomorphism on the
surface $\Sigma $, and $\frak{g}$ becomes the corresponding Lie algebra). We
can then formally (since $X\notin \frak{g}$) view the functions $X$ as a
constant $G$-gauge field on $\mathbb{R}^{4}$, and we can considers the
objects $U(y)$ as defining a gauge transformation. We may then analyze the
gauge-transformed potential 
\begin{equation}
A_{a}(y)=UX_{a}U^{-1}+\frac{i}{2\pi }U\partial _{a}U^{-1}.  \label{eq43}
\end{equation}
and we can show that, as opposed to the objects $X_{a}$ and $U(y)$, the $%
A_{a}$'s are well defined functions on the surface $\Sigma $ parametrized by
the coordinates $y$ on $\mathbb{R}^{4}$. In order to do so we first record
the following identity 
\begin{equation*}
e^{Y}Xe^{-Y}-e^{Y}e^{-Y+X}=-1+\sum_{n=1}^{\infty }d_{n}\text{Ad}^{n}(Y)(X)+\
o(X^{2}),
\end{equation*}
where 
\begin{equation*}
d_{n}=\left( \frac{1}{n!}-\frac{1}{(n+1)!}\right) .\,\,\ \ \ \ \ \ \ \ \ \ 
\begin{array}{ccc}
d_{1}=\frac{1}{2} & d_{2}=\frac{1}{3} & \cdots
\end{array}
\end{equation*}
Substituting $X\rightarrow X_{a}=-\frac{i}{2\pi }\partial _{a}(2\pi
i\,y\cdot X)$ and $Y\rightarrow 2\pi i\,y\cdot X$ (note that when we take
the derivative $\partial _{a}$ in equation (\ref{eq43}) we are actually
computing the linear term in $X_{a}$), we can use the above expression to
obtain 
\begin{equation}
A_{a}(y)=\sum_{n=1}^{\infty }d_{n}\,\text{Ad}^{n}(2\pi i\,y\cdot X)(X_{a})
\label{eq46}
\end{equation}
It is now apparent that the gauge fields $A_{a}$ only depend on the
commutator of the coordinate functions, and are therefore well defined
functions on the surface.

Up to now we have considered the functions $A_{a}(y)$ as $G$-gauge
potentials over the whole euclidean $4$-space $\mathbb{R}^{4}$. On the other
hand we can easily show that the gauge transformation $U(y)$ gives, in fact,
a well defined non-trivial bundle on the torus $\tilde{T}_{4}$. To prove
this fact, let us choose a point $y\in \mathbb{R}^{4}$ and an element $b\in 
\tilde{\Gamma}$. When we pass from $\mathbb{R}^{4}$ to the quotient $\tilde{T%
}_{4}=\mathbb{R}^{4}/\tilde{\Gamma}$, the fibers at $y$ and at $y+b$ are
glued using the transition function 
\begin{equation*}
U^{-1}(y)U(y+b)=e^{-2\pi i\,y\cdot X}e^{2\pi i\,(y+b)\cdot X}
\end{equation*}
Using the Campbell-Hausdorf formula we can rewrite the above as 
\begin{equation}
U^{-1}(y)U(y+b)=e^{2\pi i\,b\cdot X+\text{commutators}}  \label{eq41}
\end{equation}
and we therefore see that, although the single $U$'s are not well-defined,
the right hand side of (\ref{eq41}) is the exponential of a function defined
up $2\pi i\mathbb{Z}$, and therefore does represent a valid transition
function between fibers.

We are therefore left with a non-trivial bundle on the dual torus, with a
specific gauge potential $A$. To compute the curvature of the connection, we
first note that implicit in equation (\ref{eq43}) is the relation 
\begin{equation*}
D=d-2\pi i\,A
\end{equation*}
between the covariant derivative $D$ and the connection $A$. Therefore the
curvature $2$-form $F=[D,D]$ is given explicitly by 
\begin{equation}
F_{ab}=\partial _{a}A_{b}-\partial _{b}A_{a}-2\pi i(A_{a}\star
A_{b}-A_{b}\star A_{a})  \label{eq301}
\end{equation}
It is then immediate, starting from equations (\ref{eq42}) and applying the
gauge transformation $U$, to conclude that the curvature $F$ is
almost-anti-self-dual (AASD from now on). More precisely we have that 
\begin{eqnarray*}
F_{12}+F_{34} &=&-\pi \varepsilon \\
F_{13}-F_{24} &=&F_{14}+F_{23}=0
\end{eqnarray*}
Up to this point, the whole discussion has been in terms of the algebra $%
\frak{g}$ of real function on the surface. In fact, we have considered the
fiber bundle on $\tilde{T}_{4}$ as a principal bundle with underlying gauge
group $G$, whose Lie algebra is $\frak{g}$ itself. We may now use the
quantization scheme described in the previous sections. First of all, we
pass from the algebra $\frak{g}$ of real functions to the algebra $\frak{u}%
(N)$ of Hermitian operators acting on $\mathcal{H}$. At the same time the
group $G$ is replaced with the group $U(N)$, and the quantization of the
connection $A$ gives us a Hermitian connection 
\begin{equation*}
\mathcal{A}_{a}=\mathcal{Q}(A_{a})
\end{equation*}
with corresponding curvature 
\begin{equation*}
\mathcal{F}_{ab}=\mathcal{Q}(F_{ab})=\partial _{a}\mathcal{A}_{b}-\partial
_{b}\mathcal{A}_{a}-2\pi i\left[ \mathcal{A}_{a},\mathcal{A}_{b}\right] .
\end{equation*}
We are then left with a non-trivial $U(N)$ principal bundle over the torus $%
\tilde{T}_{4}$, with a specific connection whose curvature is AASD, in the
sense that it satisfies 
\begin{eqnarray}
\mathcal{F}_{12}+\mathcal{F}_{34} &=&-\pi \varepsilon  \label{eq401} \\
\mathcal{F}_{13}-\mathcal{F}_{24} &=&\mathcal{F}_{14}+\mathcal{F}_{23}=0. 
\notag
\end{eqnarray}

In order to study the properties of this principal bundle, it is natural to
compute the Chern classes 
\begin{eqnarray*}
c_{1} &=&\text{tr}(\mathcal{F}) \\
c_{2} &=&\frac{1}{2}\text{tr}^{2}(\mathcal{F})-\frac{1}{2}\text{tr}(\mathcal{%
F}\wedge \mathcal{F}).
\end{eqnarray*}
On the other hand, for reasons that will become clear later, before we start
the computation we need to make a digression and to consider some simple
topological properties of the surface $\Sigma $ embedded in $T_{4}$. Let us
first introduce, on the torus $T_{4}$, coordinates $t_{A}$ defined by 
\begin{eqnarray*}
t_{A}a_{a}^{A} &=&x_{a} \\
t_{A} &=&x_{a}b_{A}^{a}.
\end{eqnarray*}
These coordinates run from $0$ to $1$ and it is very natural to use them in
any topological descriptions of the torus $T_{4}$. In particular the second
integral de-Rahm cohomology group $H_{\text{de-Rahm}}^{2}(T_{4})$ is
generated by the forms 
\begin{equation*}
\alpha _{AB}=dt_{A}\wedge dt_{B}.\,\ \ \ (A<B)
\end{equation*}
In a similar way the second integral homology group $H_{2}(T_{4})$ is
generated by the Poincar\'{e} duals of the forms $\alpha _{AB}$. These are
the simplicies 
\begin{equation*}
\Delta ^{AB}\,\ \ \ \ \ (A<B)
\end{equation*}
extending in the $AB$ direction and satisfying 
\begin{equation}
\int_{\Delta ^{CD}}\alpha _{AB}=\delta _{A}^{C}\,\delta _{B}^{D}-\delta
_{A}^{D}\,\delta _{B}^{C}.  \label{eq44}
\end{equation}
Let us now consider $\rho (\Sigma )$ as an element of $H_{2}(T_{4})$. We can
write 
\begin{equation*}
\rho (\Sigma )=\frac{1}{2}C_{AB}\Delta ^{AB},
\end{equation*}
where the coefficients $C_{AB}$ are integral, and the equality should be
understood in the sense of homology. The $C$'s can be computed by using the
formula (\ref{eq44})\ and by noting that 
\begin{equation*}
\int_{\Sigma }\rho ^{\ast }(\alpha _{AB})=\int_{\rho (\Sigma )}\alpha
_{AB}=\int_{\frac{1}{2}C_{CD}\Delta ^{CD}}\alpha _{AB}=C_{AB}.
\end{equation*}
Using the fact that $\rho ^{\ast }(dx_{a})=dX_{a}$ and the fact that $\rho
^{\ast }(a\wedge b)=\rho ^{\ast }(a)\wedge \rho ^{\ast }(b)$, we conclude
that 
\begin{eqnarray*}
C_{AB} &=&b_{A}^{a}b_{B}^{b}\,I_{ab} \\
I_{ab} &=&\int_{\Sigma }dX_{a}\wedge dX_{b}.
\end{eqnarray*}
The coefficients $I_{ab}$ satisfy an important relation. If we consider the
K\"{a}hler form $\mu $ as a symplectic form and we define on the surface $%
\Sigma $ the corresponding Poisson bracket $\{A,B\}$ by 
\begin{equation*}
\{A,B\}\mu =dA\wedge dB,
\end{equation*}
it is the work of a moment to show that $\{Z_{1},Z_{2}\}=0$ and that $%
\sum_{i}\{Z_{i},\overline{Z}_{i}\}=-2i$. In terms of the Cartesian
coordinates these relations read 
\begin{eqnarray}
\{X_{1},X_{2}\}+\{X_{3},X_{4}\} &=&1  \label{eq48} \\
\{X_{1},X_{3}\}-\{X_{2},X_{4}\} &=&\{X_{1},X_{4}\}+\{X_{2},X_{3}\}=0.  \notag
\end{eqnarray}
If we integrate the above equations on the hole surface against the
symplectic form, and we use equation (\ref{eq300}), we immediately see that
the coefficients $I_{ab}$ satisfy the relations 
\begin{eqnarray}
I_{12}+I_{34} &=&A  \label{eq49} \\
I_{13}-I_{24} &=&I_{14}+I_{23}=0.  \notag
\end{eqnarray}

We have now all the elements needed for the computation of the Chern
classes. We start with a general remark. It is a well known fact that the
classes $c_{1}$ and $c_{2}$ are integral classes. On the other hand we see
that, given the expression (\ref{eq46}) for the gauge potential and the
formulae (\ref{eq101},\ref{eq34}) for the star product and the trace, we
expect to be able to write both $c_{1}$ and $c_{2}$ as power series in the
quantization parameter $\varepsilon $, which we recall takes discrete values 
$\varepsilon =A/\pi n$. This means that any contribution to the Chern
classes with a positive power of $\varepsilon $ must vanish in cohomology,
and therefore we might as well compute the Chern classes in the limit $%
\varepsilon \rightarrow 0$. In this limit we first of all notice that 
\begin{equation*}
\lbrack ,]\rightarrow \frac{\varepsilon }{2i}\{,\}.
\end{equation*}
This fact can then be used to simplify both the expression (\ref{eq46}) for
the gauge potential 
\begin{equation}
A_{a}(y)\simeq \pi \frac{\varepsilon }{2}y^{b}\{X_{b},X_{a}\}  \label{eq500}
\end{equation}
and the formula (\ref{eq301}) for the curvature 
\begin{equation}
F_{ab}=\pi \varepsilon \{X_{a},X_{b}\}.  \label{eq302}
\end{equation}
Finally, recalling the result (\ref{eq33}) regarding traces of operators in
the $\varepsilon \rightarrow 0$ limit, we conclude that 
\begin{eqnarray*}
c_{1} &=&\text{tr}(\mathcal{F})=\frac{1}{2}dy^{a}\wedge dy^{b}\text{tr}(%
\mathcal{F}_{ab}) = \frac{1}{2}dy^{a}\wedge dy^{b}\int_{\Sigma }\{X_{a},X_{b}\}\mu = \\
&=&\frac{1}{2}dy^{a}\wedge dy^{b}\int_{\Sigma }dX_{a}\wedge dX_{b}=\frac{1}{2%
}I_{ab}dy^{a}\wedge dy^{b}.
\end{eqnarray*}
First of all we notice that all of the $\varepsilon $ dependance has
vanished, as we expected. To show that the above expression actually does
represent an integral form, we introduce, like in the case of the torus $%
T_{4}$, coordinates $s_{B}$ on $\tilde{T}_{4}$ defined by 
\begin{equation*}
s^{B}b_{B}^{b}=y^{b}.
\end{equation*}
The integral cohomology $H_{\text{de-Rahm}}^{2}(\tilde{T}_{4})$ is then
generated by the forms 
\begin{equation*}
\beta ^{AB}=ds^{A}\wedge ds^{B}.
\end{equation*}
In terms of these coordinates, we can write that$^{{}}$%
\begin{eqnarray*}
c_{1} &=&\frac{1}{2}I_{ab}dy^{a}\wedge dy^{b}= \\
&=&\frac{1}{2}I_{ab}b_{A}^{a}b_{B}^{b}ds^{A}\wedge ds^{B}=\frac{1}{2}%
C_{AB}\beta ^{AB},
\end{eqnarray*}
thus proving that $c_{1}$ is an integral class, as expected.

We now move to the second Chern class. In this case the computation is much
simpler for the following reason. We have seen that the curvature $\mathcal{F%
}$ is of order $\varepsilon $ so that $\mathcal{F}\wedge \mathcal{F}\sim
\varepsilon ^{2}$. Traces, on the other hand, can be considered to be of
order $\varepsilon ^{-1}$, so that tr$(\mathcal{F}\wedge \mathcal{F})\sim
\varepsilon $. By the previous reasoning we then expect tr$(\mathcal{F}%
\wedge \mathcal{F})$ to vanish in cohomology, and therefore we conclude that 
\begin{equation*}
c_{2}=\frac{1}{2}c_{1}^{2}.
\end{equation*}
To convince ourselves of this result, we may check that, to lowest order, tr$%
(\mathcal{F}\wedge \mathcal{F})$ does indeed vanish. Using the expression (%
\ref{eq302}) for the curvature, and again using the formula (\ref{eq33}) for
the trace, we see that 
\begin{eqnarray*}
\text{tr}(\mathcal{F}\wedge \mathcal{F}) &\sim &\,dy^{a}\wedge dy^{b}\wedge
dy^{c}\wedge dy^{d}\int_{\Sigma }\,\mu \{X_{a},X_{b}\}\{X_{c},X_{d}\} \\
&\sim &\,d^{4}y\int_{\Sigma }\,\mu
(2\{X_{1},X_{2}\}\{X_{3},X_{4}\}-2\{X_{1},X_{3}\}\{X_{2},X_{4}\} \\
&&+2\{X_{1},X_{4}\}\{X_{2},X_{3}\})
\end{eqnarray*}
Using equations (\ref{eq48}) we can rewrite the above as 
\begin{equation*}
d^{4}y\left( A-\frac{1}{2}\int_{\Sigma }\,\mu \{X_{a},X_{b}\}^{2}\right)
\end{equation*}
To show that this final expression vanishes, we just need to show that $%
\frac{1}{2}\{X_{a},X_{b}\}^{2}=1$. On the other hand, recalling that $%
\{A,B\}=Q^{-1}\varepsilon ^{\alpha \beta }\partial _{\alpha }A\partial
_{\beta }B$, we can easily check that $\det \,h_{\alpha \beta }=\frac{1}{2}%
Q^{2}\{X_{a},X_{b}\}^{2}$, where $h_{\alpha \beta }=\partial _{\alpha
}X_{a}\partial _{\beta }X_{a}$ is the induced metric on the surface. But
since $\mu =\sqrt{\det \,h_{\alpha \beta }}dxdy=Qdxdy$, we have that $\det
\,h_{\alpha \beta }=Q^{2}$, and therefore that 
\begin{equation*}
\frac{1}{2}\{X_{a},X_{b}\}^{2}=1.
\end{equation*}

We have therefore shown that holomorphic curves embedded in compact tori
have a matrix representation as $U(n)$ connections over the dual tori, with
AASD curvature. The underlying principal bundles are non-trivial, with
vanishing instanton number and with first Chern class corresponding to the
homology class of the surface embedded in the target space.

Let me describe briefly the simplest example of the formalism just outlined.
We will rederive in a complex way some simple known results in order to
connect the construction just described to a more familiar context. We will
take $T_{4}$ to be the unit cube, and we will let $\Sigma $ be the $g=1$
complex surface with modular parameter $\tau =i$. Let $\sigma _{1}$, $\sigma
_{2}$ be the canonical coordinates on $\Sigma $ and consider the embedding 
\begin{eqnarray*}
x_{1} &=&a\sigma _{1}\,\ \ \ x_{2}=a\sigma _{2}\,\ \ \ \ \ \ \ \ (a,b\in 
\mathbb{Z}) \\
x_{3} &=&b\sigma _{1}\,\ \ \ x_{4}=b\sigma _{2}
\end{eqnarray*}
so that 
\begin{eqnarray*}
\mu &=&(a^{2}+b^{2})d\sigma _{1}d\sigma _{2} \\
A &=&a^{2}+b^{2}.
\end{eqnarray*}
One can easily check that 
\begin{eqnarray*}
\{X_{1},X_{2}\} &=&1-\{X_{3},X_{4}\}=\frac{a^{2}}{A} \\
\{X_{1},X_{3}\} &=&\{X_{2},X_{4}\}=0 \\
\{X_{1},X_{4}\} &=&-\{X_{2},X_{3}\}=\frac{ab}{A}.
\end{eqnarray*}
This implies that (using equation (\ref{eq500}) and the fact that $%
\varepsilon =\frac{A}{\pi n}$) we may represent the embedded surface with a $%
U(n)$ Yang-Mills linear connection on the dual torus given by 
\begin{eqnarray*}
A_{1}(y) &=&\frac{1}{2n}(-y_{2}a^{2}-y_{4}ab)\,\mathbf{1}_{n\times n} \\
A_{2}(y) &=&\frac{1}{2n}(y_{1}a^{2}+y_{3}ab)\,\mathbf{1}_{n\times n} \\
A_{3}(y) &=&\frac{1}{2n}(-y_{2}ab-y_{4}b^{2})\,\mathbf{1}_{n\times n} \\
A_{4}(y) &=&\frac{1}{2n}(y_{1}ab+y_{3}b^{2})\,\mathbf{1}_{n\times n}.
\end{eqnarray*}

To conclude this section, we would like to discuss the question of stability
of the AASD configurations that we have analyzed.

Purely within the context of Yang-Mills theory, we might worry that the
solutions just described do not represent a local minimum of the YM action.
In fact, using the AASD property of $\mathcal{F}$ it is easy to show that 
\begin{eqnarray*}
S_{YM} &=&\int \text{tr}(\mathcal{F\wedge \star F})=-\int \text{tr}(\mathcal{%
F\wedge F})+\int (\pi n\varepsilon )^{2}= \\
&=&(\pi n\varepsilon )^{2}\times Vol_{T_{4}}\sim \varepsilon ^{2}.
\end{eqnarray*}
On the other hand it is clear that equations (\ref{eq401}) define a solution
to the equations of motion. Since $\mathcal{F}+\mathcal{\star F}=\varepsilon
\,\Omega $, where $\Omega $ is a covariantly constant $2$-form, one can
still use Bianchi identity $D\mathcal{F}=0$ to show that 
\begin{equation*}
D\mathcal{\star F}=0.
\end{equation*}
To resolve this puzzle let us take the following point of view. We have seen
that to each holomorphic curve $\rho :\Sigma \rightarrow T_{4}$ we assign an
element $\frac{1}{2}C_{AB}\Delta ^{AB}\in H_{2}(T_{4})$. Moreover the
coefficients $C_{AB}=b_{A}^{a}b_{B}^{b}\,I_{ab}$ are expressed themselves in
terms of the coefficients $I_{ab}$, which satisfy the relations (\ref{eq49}%
). Therefore the fact that the $C$'s are integral imposes restrictions on
the possible values of the area $A$ of the embedded surface (for example, if 
$T_{4}$ is just the cube of unit volume, then $b_{A}^{a}=\delta _{A}^{a}$,
and $C=I$. But then $A=C_{12}+C_{34}$ is the sum of integer numbers, and is
itself an integer). We see that the requirement of holomorphic embedding,
together with the trivial fact that the homology class of $\Sigma $ is
integral, impose restrictions on the geometrical value of the area. We now
consider the situation, in some sense dual to the one just described, of a
principal $U(n)$ bundle over the dual torus $\tilde{T}_{4}$, with prescribed
integral first Chern class $c_{1}=\frac{1}{2}C_{AB}\beta ^{AB}\in H^{2}(%
\tilde{T}_{4})$. Suppose that we also have a gauge potential, whose
curvature $2$-form is AASD. If we write $\mathcal{F}=\frac{1}{2}\mathcal{F}%
_{AB}\beta ^{AB}$, it is clear that

\begin{equation*}
C_{AB}=(-1)^{A+B+1}\int_{\tilde{T}_{4}}d^{4}t\,\text{tr}(\mathcal{F}_{AB}).
\end{equation*}
Using the fact that $\mathcal{F}_{AB}=\mathcal{F}_{ab}b_{A}^{a}b_{B}^{b}$,
we can write the above as 
\begin{eqnarray*}
(-1)^{A+B}C_{AB} &=&b_{A}^{a}b_{B}^{b}\,N_{ab} \\
N_{ab} &=&-\int_{\tilde{T}_{4}}d^{4}t\,\text{tr}(\mathcal{F}_{ab}).
\end{eqnarray*}
Using the AASD property of the connection, we see that numbers $N_{ab}$
satisfy the relations 
\begin{eqnarray*}
N_{12}+N_{34} &=&\pi \varepsilon \,n \\
N_{13}-N_{24} &=&N_{14}+N_{23}=0.
\end{eqnarray*}
Like in the case discussed above, the condition of integrality of the
coefficients $C_{AB}$ imposes a constraint on the possible values of the
parameter $\varepsilon $. This result resolves the question of stability, by
proving that, purely in the context of Yang-Mills theory, the parameter $%
\varepsilon $ must be quantized. Moreover we see that the possible values of
the area and the possible values of $\varepsilon $ are related by the
formula $\pi \varepsilon \,n=A$, thus recovering in a different way the
quantization condition for $\varepsilon $.

\section{\label{c}Conclusion}

In this paper we analyze the problem of the matrix representation of
holomorphic curves embedded in complex tori. To each analytically embedded
membrane we associate a $U(N)$ Yang Mills configuration on the dual torus
which is almost-anti-self-dual. The corresponding principal bundle has
vanishing instanton number and first Chern class corresponding to the
homology class of the membrane embedded in the original torus. In order to
tackle the problem, we extend previous results on quantization of Riemann
surfaces which were derived originally in \cite{aaa}. In particular we show
that the proposed quantization scheme naturally leads to an associative star
product over the space of functions on the surface.

We conclude by suggesting some future lines of investigation

1) All of the results in this paper have been derived starting from the
asymptotic expansion in powers of $\varepsilon $ for the Bergman projection
and for the star product. On the other hand it would be desirable to
understand the non-perturbative aspects of the solution. In this context,
various directions of investigation are possible. On one hand one can try to
attach directly the problem of existence of an exact deformation $%
R(\varepsilon )$, whose asymptotic expansion is given by equation (\ref{eq6}%
), which preserves the commutators (\ref{eq42}). On the other hand it might
be more instructive to proceed in two steps. Let us suppose, as is plausible
to do, that the expansion (\ref{eq6}) is the perturbation expansion of some
auxiliary field theory living on the surface $\Sigma $. One could then
rephrase the problem of finding the non-perturbative contributions to $%
R(\varepsilon )$ in terms of the analysis of the non-perturbative structure
of the auxiliary field theory itself.

2) In this paper we have looked at the matrix representation of a fixed
holomorphic curve of genus $g$. One may then speculate about the possible
relations between the space of holomorphic curves on $T_{4}$ and the space
of AASD connections on the dual torus $\tilde{T}_{4}$. It is known that the
space of deformations of a holomorphic curve has dimension of order $g$ (in
what follows I will just do an order of magnitude discussion without keeping
track of constants of order one). One the other hand one can consider the
space of $U(N)$ AASD connections at fixed $N$. The dimension of this space
should be computable, and should be of order $N$. We recall that we can
construct, given an embedded surface, an AASD connection with $N=N(n)$,
where $N(n)=n$ for $n>g-1$. On the other hand if $g\gg n$, one can show that 
$N(n)=0$. We therefore see that, for fixed $N$, we can use surfaces of genus
up to $g\sim N$. We then see that, for surfaces of maximal genus, the
dimensions of the space of holomorphic curves and of AASD connections agree.
This gives hope that the relation between the two spaces just described can
be made sharp.

3) Finally one can hope to extend the results on Bergman projections and on
star-products to the case of manifolds of complex dimension greater then $1$%
. The main feature that should be retained is the geometric character of
expressions (\ref{eq102}) and (\ref{eq101}). In fact in the $\dim _{\mathbb{C%
}}\Sigma =1$ case equations (\ref{eq102},\ref{eq101}) are considerably
simpler then the corresponding power expansions in $\varepsilon $, as was
noted in section \ref{a}.

\section{\label{aa}Acknowledgments}

We would like to thank W. Taylor, I. Singer and V. Guillemin for very useful
discussions.

\end{document}